\documentclass{resonance}
\usepackage{graphicx}
\usepackage{float}
\usepackage{amsmath}

\begin{document}

\title{Gravitational Waves from merging binaries}
\author{Jahanvi, Ashish Kumar Meena $\&$ J. S. Bagla}

\maketitle
\authorIntro{\includegraphics[width=2.4cm]{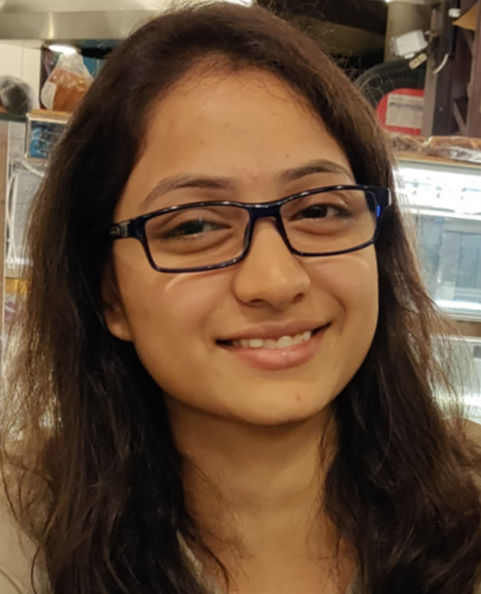} \\
Jahanvi is an MS student at IISER Mohali. Her research interest is in
Black-hole astronomy. \\
\includegraphics[width=2.4cm]{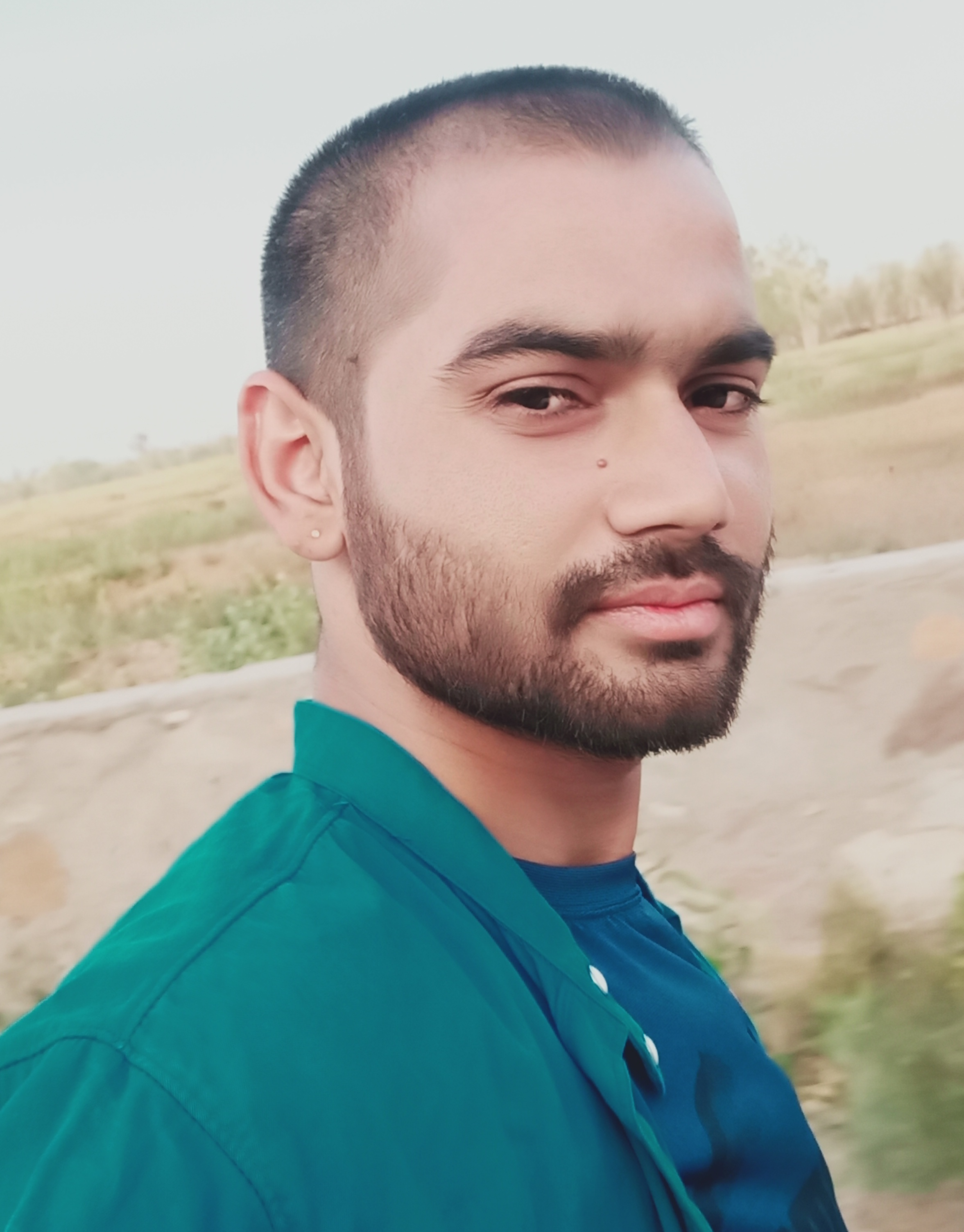} \\
Ashish is a research scholar at IISER Mohali. His research interest is
in gravitational lensing. \\
\includegraphics[width=2.4cm]{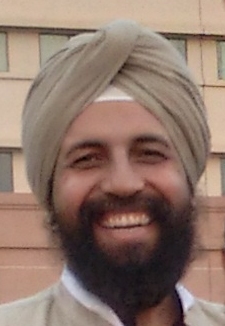} \\
Jasjeet works at IISER Mohali. He is interested in diverse problems in
physics, his research is in cosmology and galaxy formation.
}

\begin{abstract}
\textbf{
We discuss gravitational waves from merging binaries using a Newtonian
approach with some inputs from the Post-Newtonian formalism.
We show that it is possible to understand the key features of the signal
using fundamental physics and also demonstrate that an approximate
calculation gives us the correct order of magnitude estimate of the
parameters describing the merging binary system.
We build on this analysis to understand the range for different types
of sources for given detector sensitivity.
We also consider known binary pulsar systems and discuss the expected
gravitational wave signal from these.
}
\end{abstract}

\monthyear{June 2019}
\artNature{GENERAL ARTICLE}

\section{Introduction}
\label{sec:introduction}

Einstein introduced the general theory of relativity about $100$ years
ago, and since then, scientists have validated its applicability in
many ways via different experiments and
observations~\cite{2010AmJPh..78.1240W,weinberg_1972}.
One of these tests is the loss of energy due to the emission of gravitational
waves in a binary system.
This was observed in the Hulse-Taylor binary by measuring
the change in the semi-major axis and other orbital parameters over
several decades~\cite{taylor_1993, weisberg_2005}.
The gravitational waves distort the space-time curvature as they
propagate. However, the amplitude of these distortions is very
small~\cite{biman_2016, thorne_1997, schutz_2009, maggiore_2008}.
Hence the direct detection of gravitational waves was a major challenge for astronomers.
Extremely sensitive detectors are required to detect gravitational
waves and it took nearly four decades from initial concept to
detection as considerable work had to be done in order to improve the
sensitivity of the detectors~\cite{weiss_2018, barish_2018, thorne_2018}.

The effort for direct detection of these gravitational waves started
more than fifty years ago~\cite{weber_1969,gwinterfero,pressthorne}.
The gravitational waves were detected for the first time in $2015$ by
LIGO (Laser Interferometer Gravitational-Wave
Observatory)~\cite{ligo_2018} from a merging black hole binary system.
Up to second observing run (O2), a total of $11$ gravitational wave
signals have been detected by LIGO.
Out of these eleven signal, ten were from the merger of binary black
holes, and one from a binary neutron star merger.
The estimated mass of these black holes lies in a range $5$~M$_\odot$
to $60$~M$_\odot$.

With the increase in sensitivity of LIGO in the ongoing run and new
detectors (KAGRA, INDIGO, LISA)~\cite{kagra_2018, tarun_2016, indigo_2011, lisa_2012}, the
number of observed gravitational wave signals are expected to increase
significantly.
At the same time, we expect a more precise measurement of the source
parameters for many sources.
The detections of the gravitational wave signal provide a strong test
for theories of gravitation in the regime of strong gravitational
fields.
More importantly, these detections open a new window for observing
some of the most extreme events in the Universe.
Among other things, gravitational wave astronomy is expected to help
in probing the internal structure of neutron stars.
Some of the events can also be used to study the expansion of the
Universe~\cite{ajith_2011}.

In the current analysis, we discuss the possibility of detections
of different mergers by LIGO/VIRGO or LISA (Laser Interferometer Space
Antenna).
We also discuss the prospects of detecting gravitational waves from
known binary systems: Hulse-Taylor, and PSR
J0737-3039~\cite{kramer_2008}.
It turns out that at present PSR J0737-3039 is emitting gravitational
waves in LISA frequency band whereas the value of the gravitational wave frequency corresponding to Hulse-Taylor binary is smaller than the
LISA range.
As this article is written from a pedagogical perspective, in order
to calculate the required quantities, we use Newtonian approximation
with some inputs from post-Newtonian corrections.
The idea is to get an order of magnitude estimates while avoiding
complications as much as possible. 
We also assume that the binary components are spin-less.

In \S{2}, we present the mathematical formulae required for our
current analysis.
\S{3} contains a discussion of different types of mergers and
the possibility of their detection.
\S{4} includes a brief discussion of known binary systems.
We summarize the discussion of gravitational waves from binary mergers
in \S{5}.

\keywords{Gravitational waves, Binary mergers, LIGO, LISA }

\section{Gravitational waves emission}
\label{sec:math}

In this section, we set out our approach for relating the source parameters
with the gravitational wave signal.

\subsection{Keplerian Motion of a Binary System}

The energy carried away by gravitational waves as a fraction of the
total energy of the system is significant only at the time of the merger.
In such a scenario, we can use Newton's laws and Keplerian orbits to
give an approximate description of the relation between the time
period and the semi-major axis.
This is a good approximation before
the merger (see figure 1 and 2 for an illustration), the only thing
that this does not capture is the evolution 
of the orbital parameters due to the emission of gravitational waves.
Kepler's third law gives the period of the orbital motion $T$,
as~\cite{gregory_1998}
\begin{equation}
T^2 =\frac{4\pi^2}{G}\frac{a^3}{(m_1+m_2)},
\label{eq:time_period}
\end{equation}
where $a$ is the semi-major axis of the orbit and
$\left(m_1,m_2\right)$ are the masses of the two components of the
binary system.

The orbital frequency corresponding to the binary is given by the formula,
\begin{equation}
f_{orb} = \frac{1}{T} = \frac{1}{2\pi} \sqrt{\frac{G (m_1+m_2)}{a^3}}
\label{eq:freq_orbital}
\end{equation}
Since, the frequency of emitted gravitational waves is twice the
orbital frequency, the frequency of the outgoing gravitational waves
is given by, ~\cite{carroll_2004}
\begin{equation}
f_{gw} = 2 f_{orb} = \frac{1}{\pi} \sqrt{\frac{G (m_1+m_2)}{a^3}}.
\label{eq:freq_initial}
\end{equation}
In the weak field limit, it can be shown that the gravitational waves
propagate at the speed of light, are transverse in nature and like the
electromagnetic waves these have two polarizations \cite{biman_2016}.
For binaries in circular orbits, the strain due to the two
gravitational wave polarizations at a distance $R$ from the binary, under
Newtonian approximation, is given as,
\begin{eqnarray}
h_{+}\left(t\right) &=& -\frac{1}{R} \frac{4G^2}{c^4} \frac{m_1 m_2}{a}
\frac{(1+\cos^2\textit{i})}{2} \:
\cos[2\pi f_{gw}\left(t-R/c\right)+ 2 \phi_0],
\nonumber \\
h_{\times}\left(t\right) &=& -\frac{1}{R} \frac{4G^2}{c^4} \frac{m_1
m_2}{a}\: \cos\textit{i} \:
\sin[2\pi f_{gw}\left(t-R/c\right) + 2 \phi_0],
\label{eq:newton_waveform}
\end{eqnarray}
where $a$ is the radius of the orbit, $i$ is the angle
between the orbital angular momentum vector of the binary and the
observer's line of sight, and $\phi_0$ is the initial phase at time
$t=0$.


\begin{figure}[!t]
\caption{Time vs. frequency plot for a $1M_\odot + 1M_\odot$ binary:
The solid black line represents the frequency evolution of gravitational
waves using equation~\eqref{eq:gw_freq}. The dashed-dotted line is the
frequency evolution under the Taylor-T4 approximant. The time $t=0$
corresponds to the stage where the semi-major axis
is $250$~km.}

\label{fig:freq_time}
\centering
\includegraphics[width=9cm]{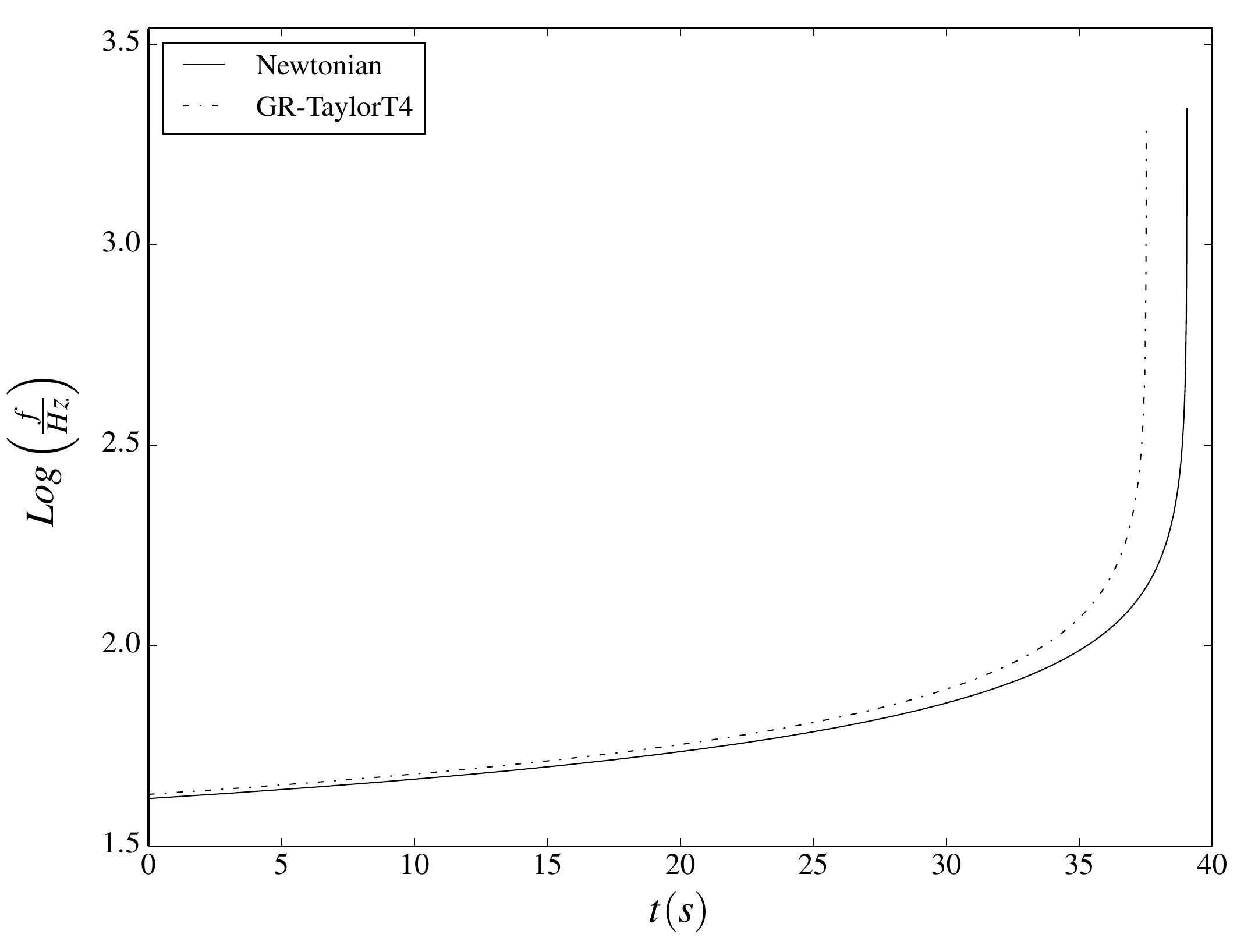}
\end{figure}


Clearly, there is a loss of energy due to the emission of gravitational
waves but this is not taken into account in our discussion of orbital
parameters.
At early times the orbital parameters vary at a very slow rate and the
above relations can be applied over multiple orbits.
At late stages, the change in orbital parameters is relatively rapid, and
these relations should be thought of as instantaneous.
It is noteworthy that strain is proportional to $1/R$. Hence, if our
sensitivity for strain measurement improves by a factor, the range up
to which we can observe events also increases by the same factor. 

In order to establish that this approximation is reasonable for the
quantities of interest in our discussion, we give the formulae for
change of orbital parameters from leading order corrections of
post-Newtonian (PN) calculations and
provide a sample calculation.

\subsection{Effects of Gravitational Wave emission}

\begin{figure}[!t]
\caption{Time vs. fractional change in frequency of outgoing
gravitational waves for a $1M_\odot + 1M_\odot$ binary under
Newtonian approximation
(corresponding to figure ~\ref{fig:freq_time}).}
\label{fig:delta_freq}
\centering
\includegraphics[width=9cm]{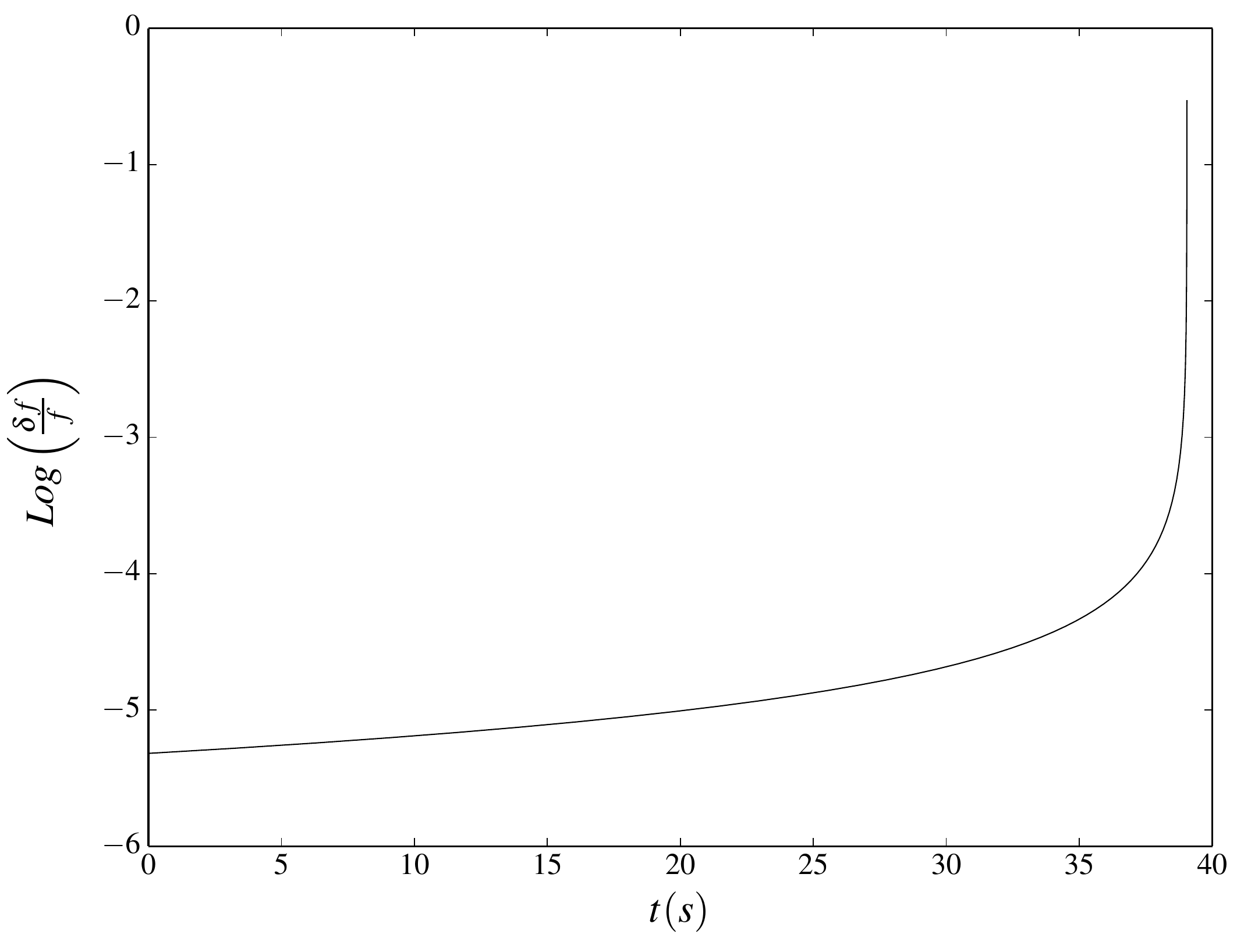}
\end{figure}

If the energy and angular momentum carried away by the gravitational
waves within the observation period are significant, then orbital
parameters and hence the frequency and amplitude of the emitted waves
vary within a few orbital time periods.
For a binary with elliptical orbit, the average rate of change of the
semi-major axis $\left(a\right)$ and eccentricity $\left(e\right)$
with the time is given as~\cite{peters_1964}
\begin{eqnarray}
\left\langle \frac{da}{dt} \right\rangle &=& -\frac {64}{5} \frac{G^{3}
m_{1} m_{2} (m_{1}+ m_{2})} {c^{5} a^{3}
(1-e^{2})^{7/2}} \left(1+
\frac {73}{24} e^{2} + \frac {37} {96} e^{4}\right),
\label{eq:da_dt} \\
\left\langle \frac{de}{dt} \right\rangle &=& -\frac {304}{15}
\frac{G^{3} m_{1} m_{2}
(m_{1}+ m_{2})}{c^{5}
a^{4} (1-e^{2})^{5/2}} e
\left(1+ \frac {121}{304}
e^{2}\right).
\label{eq:de_dt}
\end{eqnarray}
As one can see from equation \eqref{eq:da_dt} and \eqref{eq:de_dt}, as
the binary system evolves in time, the eccentricity of the orbit
decreases at a faster rate than the semi-major axis.
However, if an orbit is already circular, it remains circular.
Hence from this point onwards, until mentioned, we will be focusing
on circular orbits for binaries in order to simplify the treatment and
eliminate one parameter from the discussion.
The change in orbital parameters leads to a corresponding change in
the amplitude and frequency of outgoing gravitational waves.
Up to leading order, the variation of frequency of outgoing
gravitational waves emitted is given as,
\begin{equation}
f_{gw}(t) = \frac{5^{3/8} c^{15/8}} {8 \pi G^{5/8}} \frac
{1}{\mathcal{M}^{5/8}} (t_{c}-t)^{-3/8},
\label{eq:gw_freq}
\end{equation}
where $t_{c}$ is the coalescence time (time at which the distance
between the two binary bodies goes to zero i.e. $a(t_{c})=0$) and
$\mathcal {M}$ is the chirp mass of the binary system.
This is defined as
\begin{equation}
\mathcal{M} \equiv \frac{(m_{1}m_{2})^{3/5}} {(m_{1}+m_{2})^{1/5}}.
\label{eq:chirp_mass}
\end{equation}
Equation~\eqref{eq:gw_freq} implies that up to leading order the
frequency of the outgoing gravitational waves at a time $t$ only
depends on the chirp mass of the binary.
In addition to this, at coalescence time, the frequency of outgoing
gravitational waves goes to infinity, which indicates the failure of
the leading order approximation. To get rid of this diverging behavior
of frequency, one has to take into account the PN corrections.
Figure ~\ref{fig:freq_time}, represents the frequency evolution for a
$1M_\odot + 1M_\odot$ circular binary.
At the initial time, the orbital radius of the binary was taken to be
$250$~km which is larger than the distance at the merger by
more than a factor of 10.
The solid line represents the frequency evolution under Newtonian
approximation calculated using equation~\eqref{eq:gw_freq}, whereas
the dashed-dotted line represents the frequency evolution under
general relativistic Taylor-T4 approximant \cite{boyle_2007},
including 
post-Newtonian corrections.
One can see that at the time of the merger, the frequency of emitted
gravitational waves under Newtonian and the general relativistic approximation is of the same order.
One can see from the figure~\ref{fig:delta_freq}, which represents the
fractional change of frequency with time under Newtonian approximation,
that long before the merger,
the gravitational wave frequency evolves very slowly with time.
However, as the system proceeds towards the merger phase, which lasts
for a fraction of a second, the frequency of outgoing gravitational
waves increase very rapidly.
Here the maximum frequency represented in the figure ~\ref{fig:freq_time} corresponds to
the innermost stable circular orbit (ISCO) which is defined as
$ R_{ISCO} = 6GM/c^2$~\cite{maggiore_2008}.
Looking at the figure, one can also observe that the maximum frequency of the outgoing gravitational waves lies in the LIGO frequency range.

The corresponding gravitational wave polarization functions are given
as~\cite{maggiore_2008},
\begin{align}
h_{+}(t) &=
\begin{aligned}[t]
& -\frac{4\pi^{2/3}(G \mathcal{M})^{5/3}}{c^4
R}\frac{1+\cos^2\textit{i}}{2}\:
f_{gw}^{2/3}\left(t\right) \\
& \qquad \qquad \quad \cos\left(2\pi\int_{0}^{t}
f_{GW}\left(t'\right) dt' + 2\phi_{0}\right)
\end{aligned}
\label{eq:strain_p}\\[\jot]
h_{\times}(t)&=
\begin{aligned}[t]
& -\frac{4\pi^{2/3}(G \mathcal{M})^{5/3}}{c^4 R}
f_{gw}^{2/3}\left(t\right)\cos\textit{i} \\
& \qquad \qquad \quad \sin\left(2\pi\int_{0}^{t} f_{GW}\left(t'\right)
dt' + 2\phi_{0}\right),
\end{aligned}
\label{eq:strain_c}
\end{align}
where the symbols have their usual meaning. The above formulae can
also be used for binaries at cosmological distances by replacing chirp
mass $\mathcal{M}$ to $\left(1+z\right)\mathcal{M}$ and distance $R$
to luminosity distance to the binary. The current GW detectors can only
observe binaries at redshifts less than one. Hence the effect of redshift
will be introducing the uncertainty of factor two in the relevant
quantities, at most. Keeping that in mind and for the simplicity of
analysis, we drop the redshift dependent factor in our study.


\begin{figure}[!t]
\caption{Comparison of Newtonian (equation~\eqref{eq:strain_p} and
\eqref{eq:strain_c}) and general relativistic chirp waveform: The
black line shows the chirp waveform generated by using
equation~\eqref{eq:strain_p}. The blue line is the waveform
generated by using Taylor-T4 formula. The x-axis represents the time
elapsed since the initial frequency $f = 10\:Hz$. The y-axis
represents the corresponding strain.}
\label{fig:chirp_waveform}
\centering
\includegraphics[width=9cm, height=6cm]{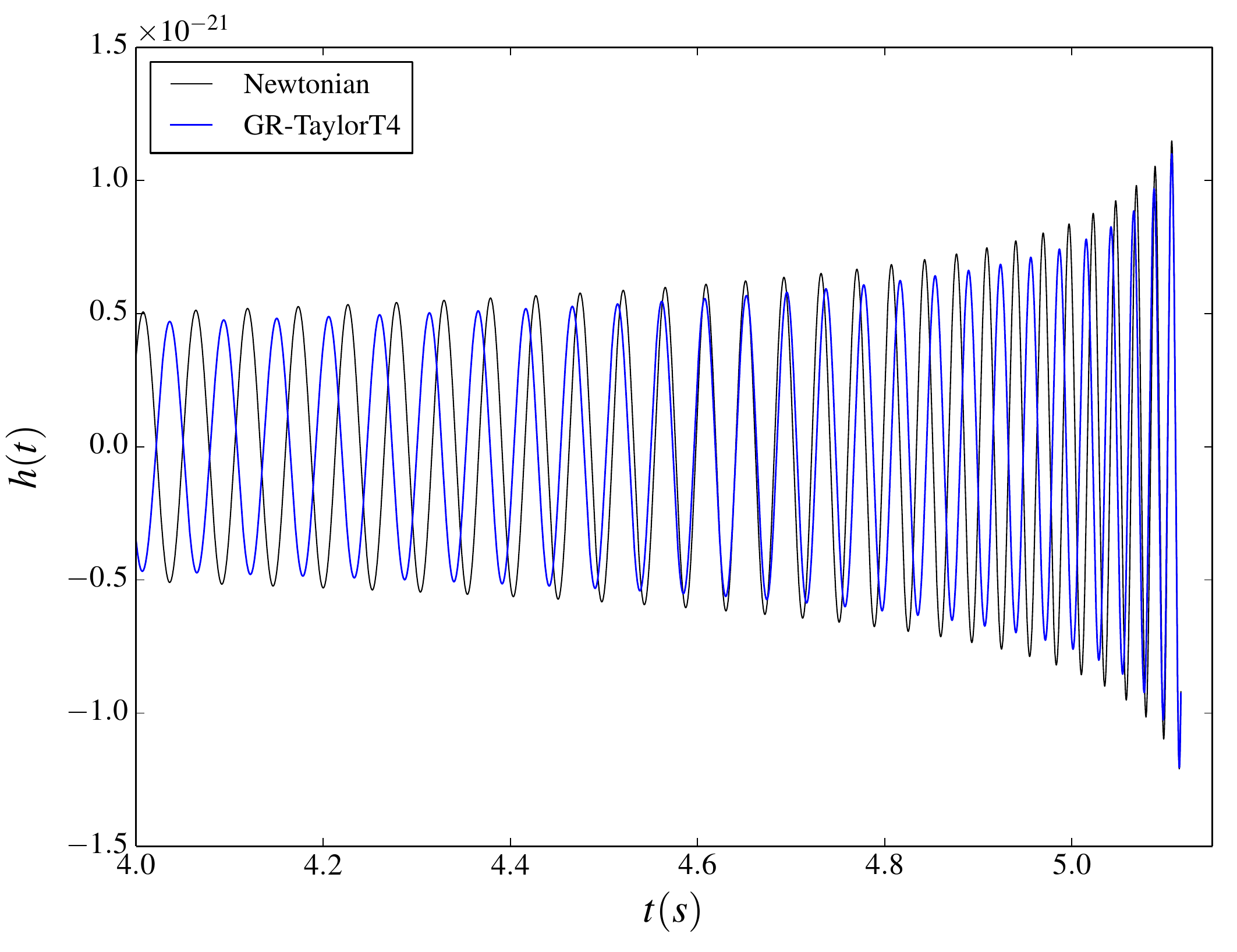}
\end{figure}


Figure~\ref{fig:chirp_waveform} shows the chirp waveforms for a
source binary which consists masses of $35.6\:M_\odot$ and
$30.6\:M_\odot$ (similar to GW150914).
The distance to the source binary is $430$ Mpc.
The continuous black line is for the waveform generated under
Newtonian approximations using equation~\eqref{eq:strain_p} whereas
the blue line represents the chirp waveform using general
relativistic (Taylor-T4) approximations. The merging time for these
two approximations is slightly different.
Here, we manually
  aligned the waveform at the time of the merger to see the change in
  the frequency and amplitude in these two approximations close to the
  time of merger as these are the two quantities used in our
  analysis. One can see that the amplitude of strain in both
  approximations remains similar. That once again validates our idea
  of using Newtonian calculations to get an estimation for frequency
  and amplitude of the GW signal. 
However, the variation of the phase of the signal with time for these
two approximations is different.
We see that the estimation of amplitude is very good and the frequency
at merger is correct to within about $10\%$.
Thus, our use of a Newtonian approximation for orbits near the time of
merger is justified.

\section{Binary Systems}
\label{sec:mergers}

Gravitational-wave detectors are sensitive to a specific frequency
bands.
Hence, the possibility of detection of the gravitational waves coming
from a source, binary depends on the frequency of gravitational waves.
The maximum frequency of gravitational waves emitted from a binary
depends on the parameters of the source binary.
The detailed waveform is used to constrain the parameters of the
source, though it is possible to get an estimate by using the
the frequency at the time of the merger.
In our discussion, we focus on the frequency at the time of the
merger, since it can be estimated without complicated calculations,
as we have shown in figure~\ref{fig:freq_time}.

In the current section, we discuss different kinds of binary
mergers which can be detected by LIGO or LISA.
Binaries can have main sequence stars (MS), white dwarfs (WD), neutron
stars (NS) or black holes (BH).
These binaries can either consist of similar kind of components
(NS-NS, BH-BH, MS-MS, WD-WD) or can be made of a different kind of
binary components (NS-BH, NS-MS, NS-WD, WD-BH, WD-MS, and BH-MS).
The prospects for detection depend on the detector sensitivity and the
frequencies where gravitational waves can be detected.
We consider different types of binaries in sub-sections below.


\begin{figure}[!t]
\caption{Frequency vs. range for same binary components: The range is
computed assuming a strain value $10^{-21}$, i.e., the detector noise is well below this level.
The light and dark gray regions represent the frequency band
corresponding to LIGO and LISA.
The horizontal lines mark some standard distances. The blue, red,
green, violet lines represent the merger of
NS-NS, BH-BH, WD-WD, MS-MS, respectively.
The large filled black circles mark objects with a mass equal to that
of the Sun.
Small filled circles mark mergers observed by LIGO, black circles are
for black holes, and the blue circle is for merging neutron stars.
See text for more details.
}
\label{fig:merger_same}
\centering
\includegraphics[width=9cm]{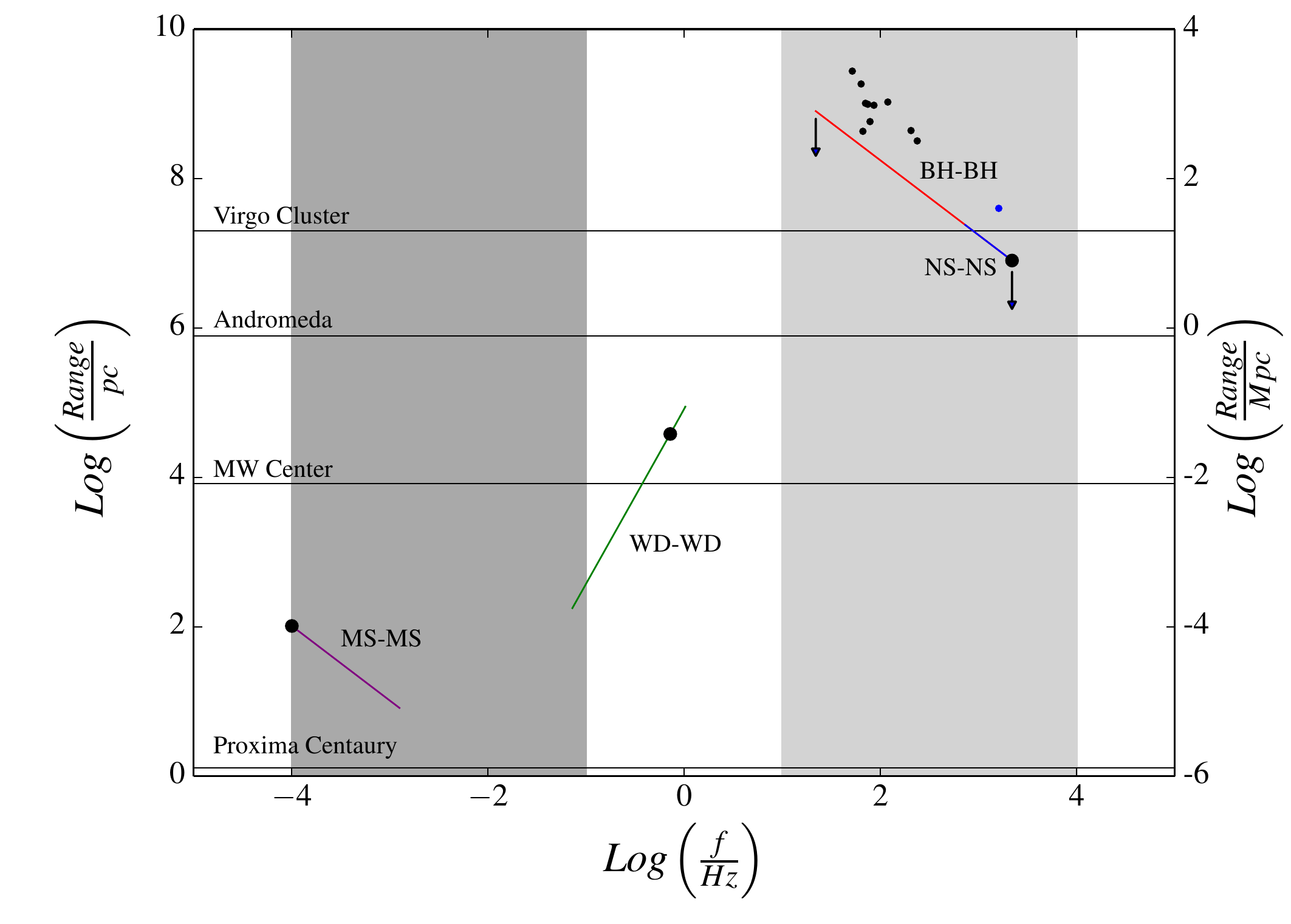}
\end{figure}


\subsection{Binaries With Similar Components}

In order to limit the number of parameters, we first consider binaries
made up of identical components.
That is, both the components are of the same type and have the same
mass.
This is adequate for our purpose, and the reader can easily generalize
this analysis using the formulae given below for any combination of
masses.  
The gravitational waves have the highest frequency when the distance
between the binary components is minimum.
For MS and WD this corresponds to the stage when the objects are
touching each other.
In case of BH and NS this corresponds to the innermost stable circular
orbit (ISCO).
$R_{ISCO}$ for an object of mass $M$ is~
\begin{equation}
R_{ISCO} = \frac{6 G M}{c^2}
\end{equation}
where symbols have their usual meaning.

For a BH-BH or NS-NS merger, this minimum distance is taken to be the
sum of $R_{ISCO}$ for each object,
\begin{equation}
a_{min} = \frac{6G\left(m_1+m_2\right)}{c^2},
\end{equation}
where $m_1$ and $m_2$ are masses of binary components.
If at some initial time $t_{in}=0$, the distance between
the binary components is $a_{in}$, then the time taken by the binary
to reach the final stage $\left(a=a_{min}\right)$ can be calculated by
integrating equation~\eqref{eq:da_dt} as,
\begin{equation}
t_{m} = \frac{a_{in}^4-a_{min}^4}{4\beta}, \,\,\,\,\,\, \beta = \frac{64}{5}\frac{G^{3}m_{1} m_{2} (m_{1}+
m_{2})}{c^{5}}
\end{equation}
From this equation, we see that the coalescence time is given as, $t_c
= a_{in}^4/4\beta$.


\begin{figure}[!t]
\caption{Frequency vs. range for same binary components in LIGO range:
The blue line represents the NS-NS mergers and red line represents
the BH-BH mergers with strain $10^{-21}$. The blue dot represents
NS-NS merger and black dots represents BH-BH mergers observed by
LIGO.
}
\label{fig:ligo_merger_same}
\centering
\includegraphics[width=9cm]{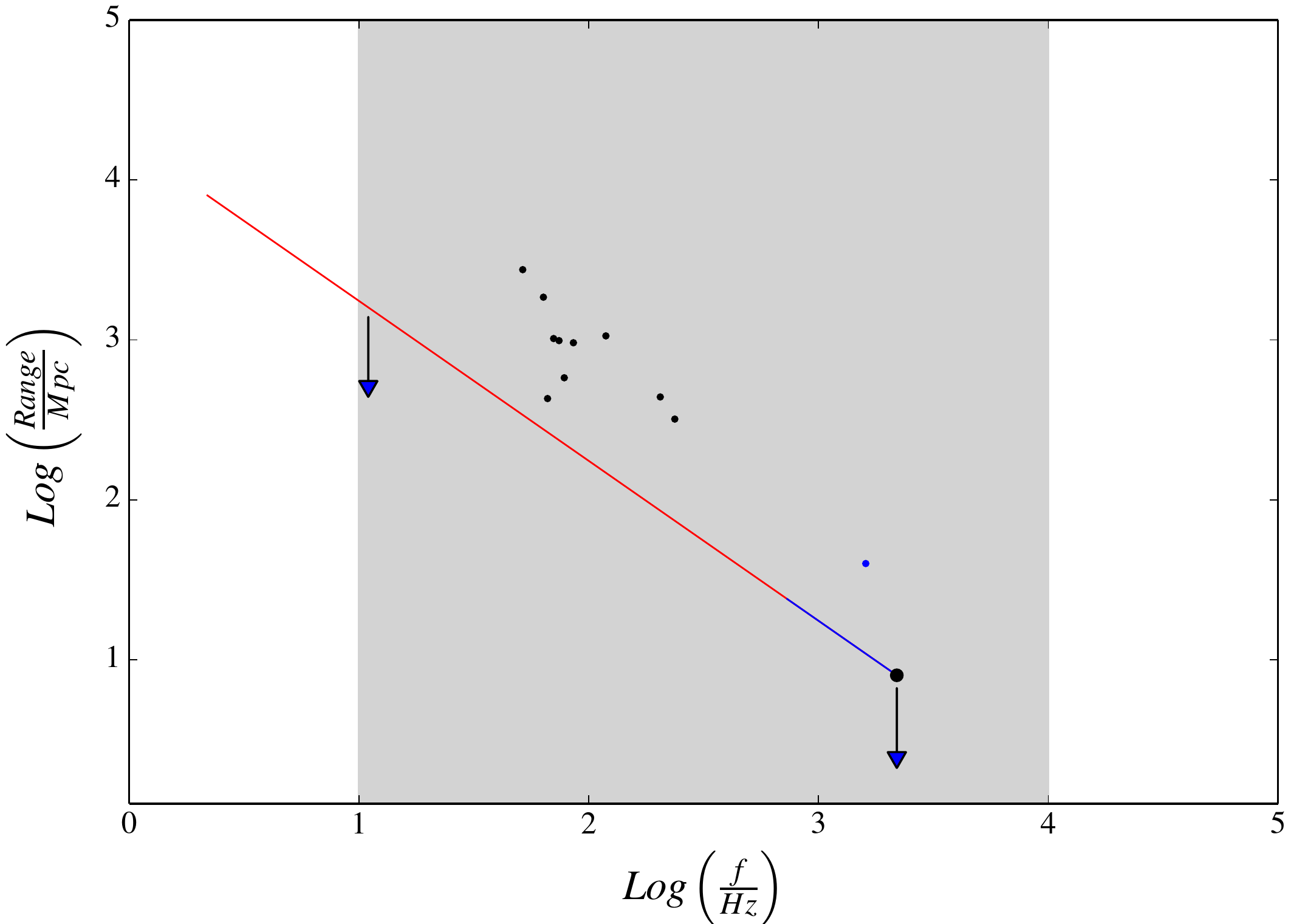}
\end{figure}


\begin{figure}
\caption{Peak frequency vs. range for different binary components:
The axis labels are similar to that of figure 4.
The light and dark gray regions represent the frequency band
corresponding to LIGO and LISA.
The horizontal lines mark some standard distances. The red, brown,
violet lines represent the merger of Black
hole with a neutron star, white dwarf and main-sequence star,
respectively. Whereas the blue and green lines represent the merger of
white dwarf and main-sequence star with a neutron star, respectively.
}
\label{fig:merger_pair}
\centering
\includegraphics[width=9cm]{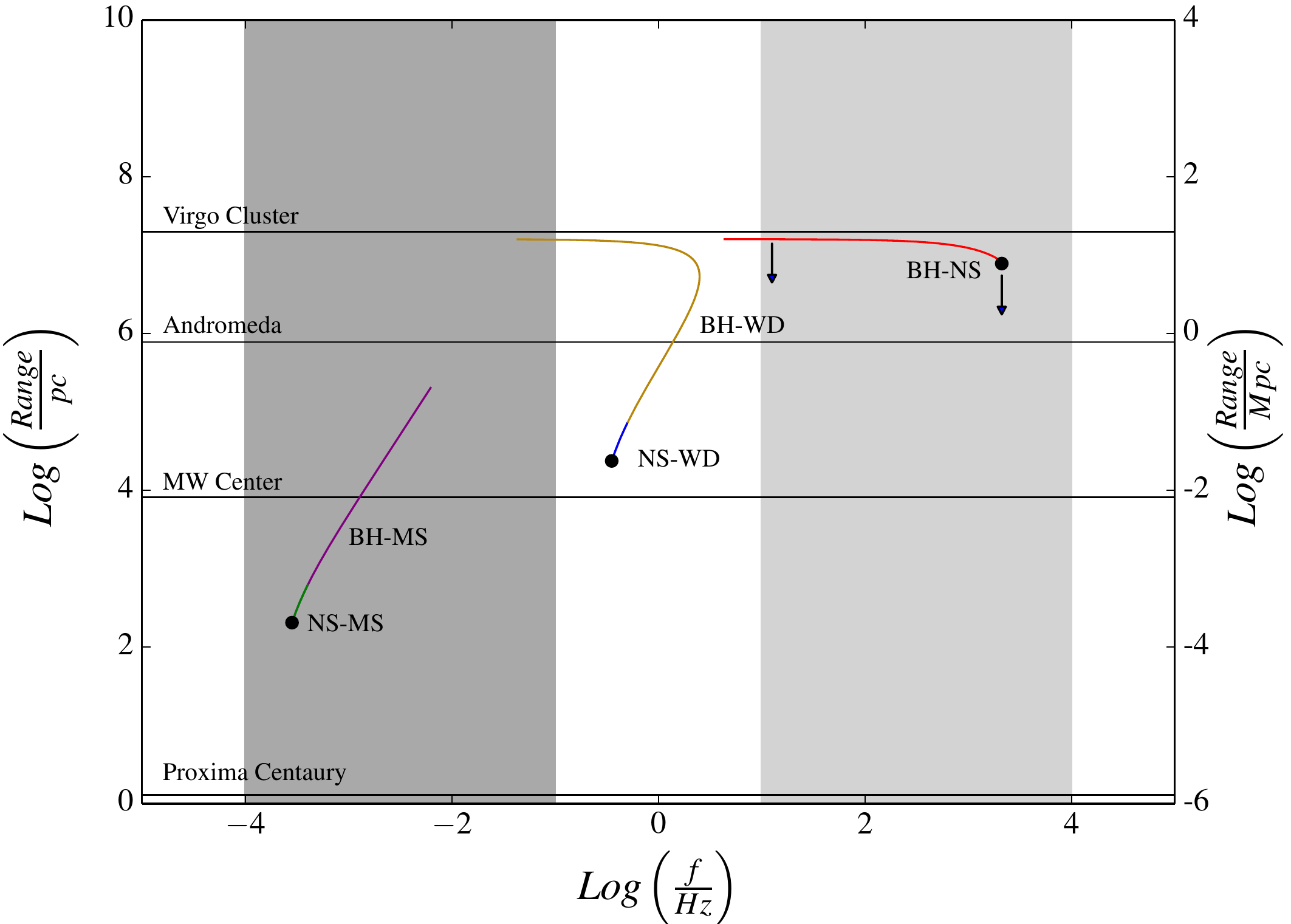}
\end{figure}


Hence from equation~\eqref{eq:freq_initial}, for a BH-BH or NS-NS merger,
the frequency of gravitational waves at $a=a_{min}$ is given as,
\begin{equation}
f_{gw}(t_{m}) = \frac{1}{\pi} \sqrt{\frac{G (m_1+m_2)}{a_{min}^3}}
\label{eq:merger_freq}
\end{equation}
The corresponding value of the range, from equation~\eqref{eq:newton_waveform},
can be given as
\begin{equation}
R = \frac{4 G^2}{c^4}\frac{m_1 m_2}{a_{min}}
\frac{1+\cos^2\textit{i}}{2}\frac{1}{h(t_{m})},
\label{eq:range}
\end{equation}
where $h\left(t_{m}\right)$, is the observed value of strain at the
merger time. We know from the equation~\eqref{eq:strain_p}
and~\eqref{eq:strain_c} that the range value is inversely
proportional to the detected strain of the GW signal. Hence, if
LIGO detects a signal with strain value $10^{-22}$ ($10^{-20}$), then
the corresponding range value will go up (down) by a factor of $10$
with respect to the range of strain value of $10^-21$. 
The sensitivity of LIGO and LISA is frequency-dependent. Hence, the
strain threshold is frequency-dependent if we work with a given signal
to noise ratio (SNR). We ignore this and fix the strain value to
$10^{-21}$ to keep the analysis simple. However, we do provide a 
qualitative indication on how this affects the range, where relevant. 
The above formulae for calculating merger frequency and range are
valid for all kinds of binaries only the values of $a_{min}$ are
different for different binaries.
Note that we are using equation~~\eqref{eq:freq_initial} and
equation~\eqref{eq:newton_waveform} for these calculations.

The mass-radius relation for a white dwarf is given
as~\cite{1939isss.book.....C,dina_2010},
\begin{equation}
R_{WD} \propto M_{WD}^{-1/3},
\label{eq:white_dwarf}
\end{equation}
where $R$ is the radius and $M$ is the total mass of the white dwarf.
This relation applies for white dwarfs with non-relativistic
degenerate electron gas but we choose to extrapolate it to the
Chandrasekhar mass limit.
In order to fix the constant of proportionality, we take the radius
for a solar mass white dwarf to be $6000$~km.
Hence the minimum distance attained by a WD-WD binary with
mass $(m_1,m_2)$ is given as
\begin{equation}
a_{min} =
6000\:\mathrm{km}\left\lbrace\left(\frac{M_\odot}{m_1}\right)^{1/3} +
\left(\frac{M_\odot}{m_2}\right)^{1/3}\right\rbrace
\end{equation}
For a main-sequence star, the mass-radius relation is given as ~\cite
{dina_2010}
\begin{equation}
R_{ms} \propto M_{ms}^{\alpha}
\end{equation}
where $\alpha = 0.8$ for a main-sequence stars with mass greater than
$1 M_\odot$ and $\alpha = 1.0$ for stars with $0.08$~M$\odot \leq M
\leq 1$~M$_\odot$.
The minimum distance attained by
an MS-MS binary with mass $\left(m_1, m_2\right)$ is given as
\begin{equation}
a_{min} = R_\odot \left(\frac{m_1^\alpha}{M_\odot} +
\frac{m_2^\alpha}{M_\odot}\right),
\end{equation}
where $R_\odot$ is the radius of the Sun.
The corresponding value of frequency and range at the time of the merger
can be calculated by using equations~\eqref{eq:merger_freq} and
\eqref{eq:range}.
Figure~\ref{fig:merger_same} shows the merger of binaries with
identical binary components with equal masses.
The dark and light-shaded regions represent the frequency range
corresponding to LISA and LIGO detectors.
The x-axis represents the frequency of emitted gravitational waves at
the time of the merger.
The y-axis represents the range up to which a gravitational wave
signal can be detected with a strain value of $10^{-21}$ by the
detector.
The LIGO detectors have a higher sensitivity so we can see mergers out
to larger distances.
Further, it has to be kept in mind that our estimates are based on
an approximate analysis.
The black horizontal lines mark some key distances.
The red line marks the merger of
binary black holes with the same masses from $3$~M$_\odot$ to
$100$~M$_\odot$.
Similarly, the blue line shows the merger of binary neutron stars
with the mass in a range from $1$~M$_\odot$ to $3$~M$_\odot$.
The black dot on the blue line represents the merger of a $1$~M$_\odot
+ 1$~M$_\odot$ neutron star binary.
The black dots above the red line show the observed BH-BH
mergers and the blue dot is for the observed neutron star
merger.
The frequencies are plotted in the source frame and not the observer
frame for ease of comparison.
One can see that LIGO can detect BH-BH as well as NS-NS mergers.
The BH binaries are the most distant binaries that can be detected by
LIGO.
However, as we know from the noise curves for the LIGO detector that at
the lower and higher end of the allowed frequency range, the noise level
is higher than the middle of the frequency range. Hence, the signal to
noise ratio (SNR) at these frequency values also goes down. As a
result, the maximum distance up to which LIGO can detect a binary
merger at these frequencies also decreases. Hence the observed range
value will be smaller than the value predicted by using
equation~\ref{eq:range}. This effect is denoted by downward arrows
in figure~\ref{fig:merger_same}.
On the other hand, the main-sequence star binaries with component mass
less than or equal to one solar mass can be detected by LISA, but the
distances up to which these binaries can be detected are much smaller
than the size of our own galaxy.
Indeed, for the assumed sensitivity, we can only detect such mergers
in the local neighborhood of the Sun within a radius of about a few
hundred parsecs.
The black dot on the violet line represents the merger of $1$~M$_\odot
+ 1$~M$_\odot$ main-sequence star binary.
Mass of components is lower for the higher frequency at merger.
Some white dwarf binaries with masses on the lower end
($0.1$~M$_\odot$) can be detected by LISA, but higher mass WD mergers
cannot be detected by LISA or LIGO.
The black dot on the green line marks the merger of a $1$~M$_\odot
+ 1$~M$_\odot$ white dwarf binary.
Mass of WD is higher for the higher frequency at the merger.

Mergers in the frequency range of LIGO are also shown in
figure~\ref{fig:ligo_merger_same} to focus on the relevant distance
scale.
The range for BH mergers is much higher than the range for NS
mergers.
This implies that we will be able to see many more BH mergers even
though the expected rate of such mergers per galaxy is much smaller.
This is because the volume within which we can see a merger of a given
type scales as the cube of the range.
We see that the distance to the detected BH mergers lies in a range
from $100$~Mpc to $3000$~Mpc whereas the distance to NS binary is
around $40$~Mpc.
If we gradually increase the mass of black holes, the maximum
frequency of gravitational waves decreases and moves towards the LISA
range.

\subsection{Binaries With Different Components}

In this section, we explore mergers of dissimilar objects.
This possibility includes the merger of BH-NS, BH-WD, BH-MS, NS-WD,
NS-MS and WD-MS.
We now use the sum of cutoff radius for each component to obtain the
smallest distance that marks the merger: for MS and WD it is the sum
of individual radii, and for NS and BH we take the radius of the
innermost stable circular orbit.
Thus for a WD-NS merger, we take the physical radius of the WD and add
$R_{ISCO}$ for the NS to obtain the semi-major axis at the time of
merger.

Figure~\ref{fig:merger_pair}, shows the prospects of detection of
such mergers.
As one can see LIGO can detect the merger of BH-NS, where we have
assumed that the neutron star has a mass equal to the Sun, and the
black hole has a mass in a range from $1-1000$~M$_\odot$.
Again the downward arrows show the decrement in SNR (and the range
value) at the lower and higher end of the LIGO frequency range.
The merger of a sun-like star and a neutron star or a black hole can
be detected by LISA but with a range that is limited to the local
group for the assumed sensitivity.
The merger of a white dwarf of one solar mass with a
neutron star cannot be detected by LIGO or LISA.
However, the merger of a white dwarf with a super-massive black hole
can be detected by LISA.

As one can see from figure~\ref{fig:merger_same},
\ref{fig:ligo_merger_same} and \ref{fig:merger_pair}, the neutron
star and black hole mergers can be detected by LIGO but as we increase
the mass of the black hole the maximum frequency of gravitational
waves emitted by binary decreases and gradually moves towards the LISA
domain.


\begin{figure}
\caption{Frequency vs. strain for observed binary systems:
The light and dark gray regions represent the frequency band
corresponding to LIGO and LISA.
The two black dots represent the value of frequency and strain at the
present time.
The system evolves towards higher frequencies and higher strain
amplitude.}
\label{fig:freq_hp}
\centering
\includegraphics[width=9cm]{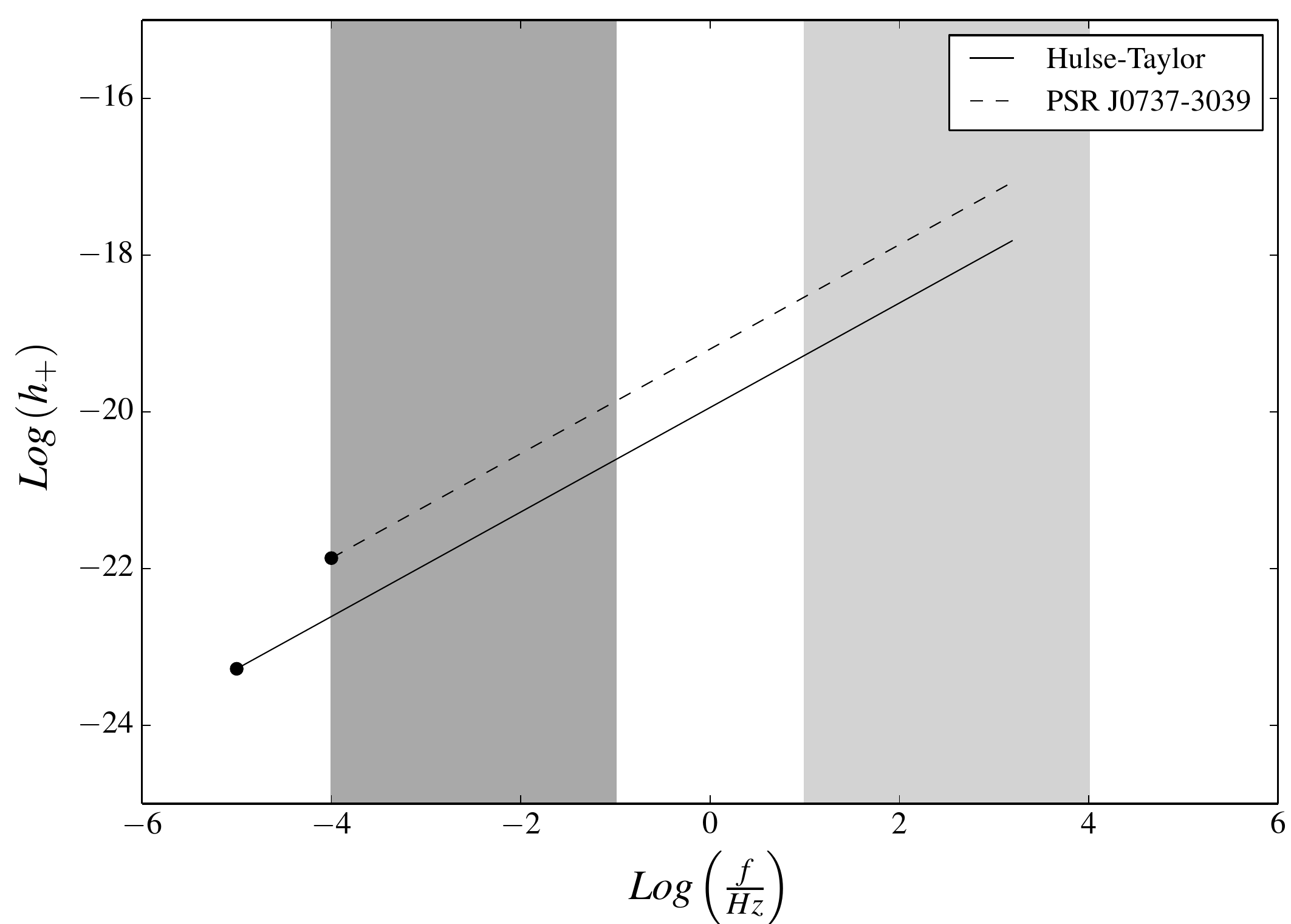}
\end{figure}


\section{Observed Binary Systems}
\label{sec:observed_system}

In this section, we briefly take a look at the known binary
systems, namely, PSR B1913+16 (Hulse-Taylor binary) and PSR
J0737-3039.
The Hulse-Taylor binary~\cite{weisberg_2005} consists of one pulsar
along with a neutron star whereas in PSR J0737-3039, both components are pulsars.

The expected merger time for the two binaries is very large:
$87$~Myr for PSR J0737-3039 and more than a Gyr for the Hulse-Taylor
binary pulsar.
Hence, the value of the frequency and amplitude of the emitted
gravitational waves are very small.
Figure~\ref{fig:freq_hp}, represents the change in frequency and strain for these
two binary systems as they proceed towards their merger phase.
The black dots represent the current value of frequency and strain for
these binaries.
As one can see, at present the frequency of outgoing gravitational
waves from PSR J0737-3039 lies in the LISA range whereas the frequency
for the Hulse-Taylor binary lies outside the LISA range.
As expected for a neutron star binary, the frequency of emitted
gravitational waves from Hulse-Taylor binary and PSR J0737-3039 at the
time of merger lies in the LIGO band.

\section{Conclusion}
\label{sec:consclusion}

We discussed gravitational waves from mergers
of different types.
We have used a simple Newtonian analysis to understand key aspects of the
expected signal.
We are able to calculate the frequency and the range at merger using a
Newtonian approach.
Given the combination of the frequency and range, it is clear that
LIGO can only detect binaries with BH and NS as its components.
LISA will be able to detect other types of mergers but the ranges for
these are relatively limited as MS and WD are less compact and the
gravitational field at the time of merger is not as strong as in the
case of BH and NS.
The range is highest for mergers of black holes, and this is reflected
in the more significant number of such events observed by LIGO so far.

It is interesting to note that LISA can detect gravitational waves
coming from PSR J0737-3039.
Observations from the square kilometer array (SKA)~\cite{ska_2013} may add to the list of
such sources.

Unlike electromagnetic radiation, gravitational waves interact very
weakly with matter, and therefore the information carried by it
undergoes fewer modifications during propagation.
As we have seen above all these signals are coming from binary
mergers, these detections can help us to understand the processes
during the merger of binary components which may not be seen
otherwise.

If one of these binary components is a neutron star, the merger leads
to the emission of electromagnetic radiation as well, as seen in GW170817.
More such sources will be discovered in the ongoing and future runs of
LIGO.
The combined analysis of gravitational waves and electromagnetic
radiation can help us in constraining the theories that describe the
structure of neutron stars and properties of matter at very high
densities, as well as cosmological parameters.

\section*{Acknowledgment}

AKM would like to thank CSIR for financial support through research
fellowship No.524007.
This research has made use of NASA’s Astrophysics Data System
Bibliographic Services.
JSB thanks members of the academic committee for the International
Olympiad for Astronomy and Astrophysics (IOAA) 2016 for their inputs:
this article was developed from a question that was developed for
IOAA-2016.

\end {document}